\documentclass[useAMS,usenatbib]{mn2e}

\usepackage{graphicx}

\title[Meteor showers of comet C/1917 F1 Mellish]
  {Meteor showers of comet C/1917 F1 Mellish}
\author[P. Vere\v{s}, L. Korno\v{s} and J. T\'{o}th]
  {P.~Vere\v{s},
  L.~Korno\v{s} and
  J.~T\'{o}th
   \\
  Faculty of Mathematics, Physics, and Informatics, Comenius University, 842 48 Bratislava, Slovakia
      Mlynsk\'{a} dolina\\
 }
\date{Released 2010 Xxxxx XX}

\pagerange{\pageref{firstpage}--\pageref{lastpage}} \pubyear{2010}

\def\LaTeX{L\kern-.36em\raise.3ex\hbox{a}\kern-.15em
    T\kern-.1667em\lower.7ex\hbox{E}\kern-.125emX}

\begin{document}

\label{firstpage}

 \maketitle

\begin{abstract}

December Monocerotids and November Orionids are weak but
established annual meteor showers active throughout November and
December. Analysis of a high quality orbits subset of the SonotaCo
video meteor database shows that the distribution of orbital
elements, geocentric velocity and also the orbital evolution of
the meteors and potential parent body may imply a common origin
for these meteors coming from the parent comet C/1917 F1 Mellish.
This is also confirmed by the physical properties and activity of
these shower meteors. An assumed release of meteoroids at the
perihelion of the comet in the past and the sky-plane radiant
distribution reveal that the December Monocerotid stream might be
younger than the November Orionids. A meteoroid transversal
component of ejection velocity at the perihelion must be larger
than 100\,m/s. A few authors have also associated December Canis
Minorids with the comet C/1917 F1 Mellish. However, we did not
find any connection.

\end{abstract}


\section{Introduction}

The comet C/1917 F1 Mellish, formerly designated as
$1917\,\,a\,\,(Mellish)$ or $Mellish\,\,1917\,\,I$, was discovered
by J.E. Mellish on March 20, 1917 and was observed for 96 days
\citep{ask23,ask32} from many places on the Earth. In the southern
hemisphere, the comet reached up to +1 magnitude. Astronomer J.F.
Skjellerup noted that the brightness of the cometary head was
about +3 magnitude, with the diffuse coma and narrow tail about
10$^\circ$ long on April 19, 1917 \citep{orch03}. The comet is a
Halley-type comet, with a relatively low inclination, and has one
of the smallest perihelion distances. It was observed only at one
apparition. Despite the relatively long observational arc, the
precision of the orbital elements is questionable. \citet{ask32}
published a slightly modified orbit of the comet and noted that
the orbit is given with a period of $145\pm0.8$\,yr
\citep{cham97}. The nominal orbital elements, according to JPL
Solar System Dynamics database \citep{cham97}, are presented in
Table 1. Although the ascending and descending node of the nominal
orbit are currently far away from the orbit of the Earth, we noted
a small difference in the eccentricity (e.g. $\Delta e\sim$
$-0.002$) would change the orbit into an Earth-crossing orbit as
the heliocentric distance of the ascending node would become equal
to 1\,AU. The orbit has a notably small perihelion distance
($0.19$\,AU). Several authors \citep{por52,has62} determined that
the comet--Earth distance is close enough to observe a meteor
shower and predicted the radiant positions and activity of the
shower on Dec. 15 (Dec. 20 respectively) and the geocentric
velocity of meteors $\sim40\,$km/s. The first few meteors
associated with the comet C/1917 F1 Mellish were obtained by the
Harvard Super Schmidt photographic survey \citep{whip54,mcc61}.
Several candidates of this meteor stream, later designated as the
December Monocerotids (MON), were also detected and distinguished
by radar surveys \citep{nil64,sek73}. Another study connecting the
December Monocerotids with the comet C/1917 F1 Mellish was made by
\citet{drum81,olson87}. Surprisingly, the radar data published by
\citet{nil64} and \citet{sek73} revealed that the meteors having
similar radiant positions, activity and geocentric velocities
appear to have 10$^\circ$ lower inclinations. \citet{kres74} noted
that December Monocerotids seem to have 2 components. The author
also speculates that the stream may have a common origin with the
Geminid meteor stream. Moreover, Harvard radio data revealed a
possible meteor stream with low inclined orbits but almost the
same orbital elements as the December Monocerotids, active between
November 27 - December 7 \citep{nil64,sek73}. The possible genetic
connection between the comet and the December Monocerotids was
studied by \citet{fox85}.

Various photographic searches confirmed the existence of a weak
stream at $RA=90.6$$^\circ$, $DEC=15.7$$^\circ$ on November 27,
with $v_{g}=43.7$\,km/s \citep{lind71}. The stream was named as
$\xi$-Orionids (Xi-Orionids, $\omega-$Orionids), currently
recognized as established meteor shower November Orionids (NOO)
within the IAU Meteor Data Center (IAU MDC) catalogue \citep{oh89}
and later by \citet{lind99}. Moreover, other photographic December
Monocerotids were published \citep{oh89} and complex analysis of
December Monocerotids and $\xi$-Orionids done by \citet{lind90}.
Even some historical records of fireballs might confirm that
December Monocerotids were active in past centuries
\citep{fox85,has99}.

In \citeyear{hind69}, \citeauthor{hind69} published his telescopic
meteor observation from Dec. 11, 1964 and assigned 5 meteors to
the new stream called 11 Canis Minorids. \citet{hind69} computed
that these meteors have parabolic orbits and much higher
inclinations (over 100$^\circ$). A year later, the author
suggested a connection between the shower and comet C/1917 F1
Mellish \citep{hind70} and determined the activity during December
9-14. \citet{kres74} revealed that 9 meteors that create the
second component of the December Monocerotids might be 11 Canis
Minorids activity within December 4-15. Their inclination was
determined as $i=29.1$$^\circ$ and the perihelion distance as
$q=0.092$, which is closer to the Sun compared to the nominal
orbit of the comet C/1917 F1 Mellish. The radiants might look like
an extension of the MON activity, however, the shower might be
active during the MON activity. The geocentric velocity is similar
$v_{g}\sim 40$\,km/s. The maximum activity of 11 Canis Minorids is
expected at $L_{\odot}=252.4$$^\circ$ (December 3).

Now, the December Monocetorids and November Orionids are weak (few
meteors per hour at maximum) but annual established meteor
showers. The shower 11 Canis Minorids, (December Canis Minorid
according to IAU nomenclature) is classified as a "working"
shower. Despite several investigations, past publications analyzed
only a small number of orbits and provided disperse data on the
mean orbit, the position of radiant, the activity and precision of
orbits, and did not reveal the orbital evolution of the meteoroid
particles released from the parent comet, in order to explain the
current state of these meteor showers. A significant number of the
analyzed orbits are hyperbolic or parabolic.

This work uses recent and precise video multi-station orbits,
obtained by the SonotaCo video network in Japan, which provide the
highest number of relatively precise meteor orbits detected
continually between the years $2007-2009$. The network operates
with over 25 similar video-optical systems and uses the same
software for meteor detection and orbit analysis (UFOCapture,
UFOAnalyser and UFOOrbit; \citealt{sonota09}). Our goal is to
provide a more complex description of meteor showers related to
the comet C/1917 F1 Mellish and reveal some of their obscured
characteristics. The eventual objective is the investigation of
the orbital evolution of meteors and the parent comet, and
determination of ejection velocities near the perihelion.

\begin{figure}
\centerline{\includegraphics[width=8cm]{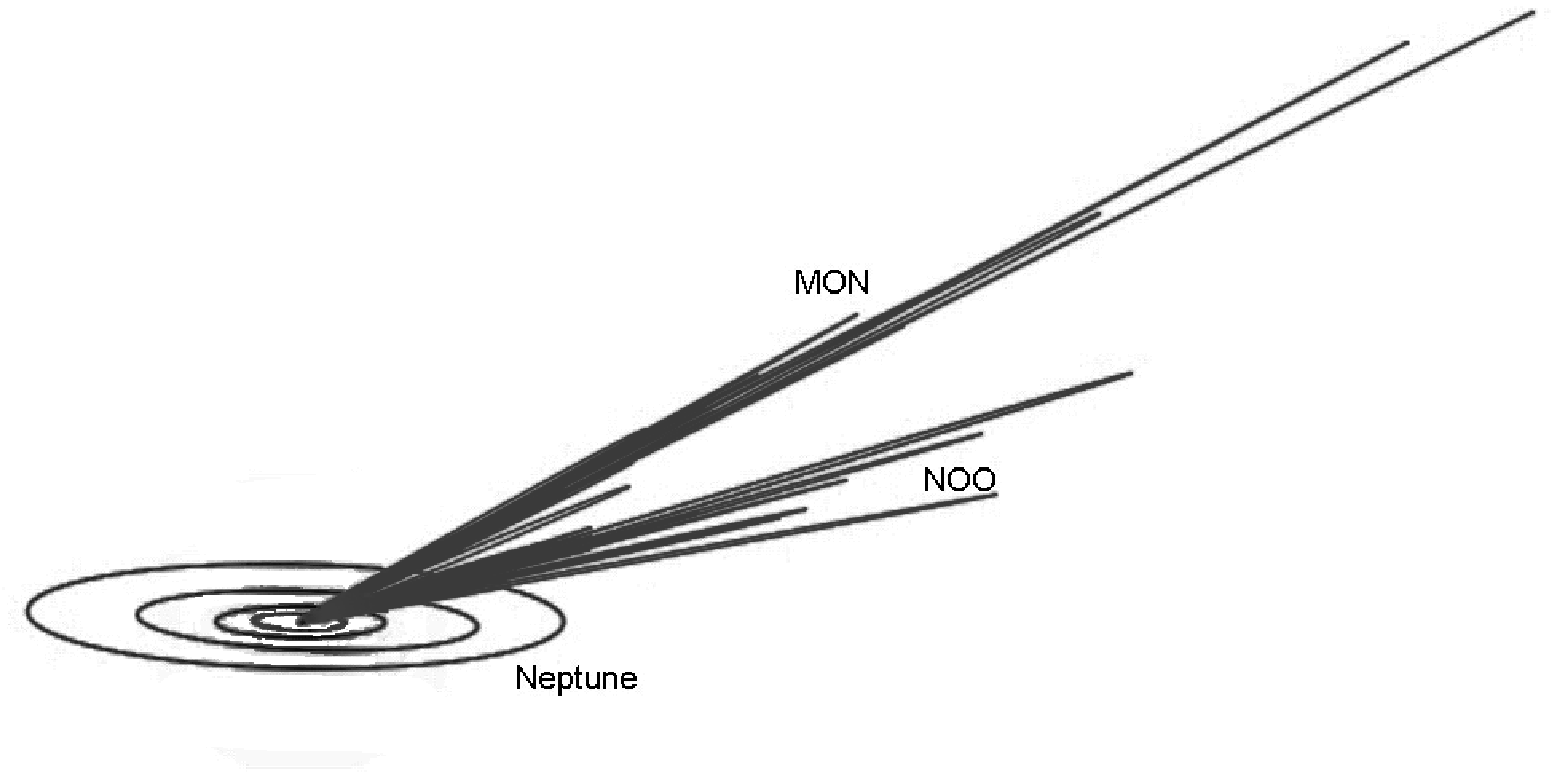}}
\vspace{-1.0cm} \centerline{
\includegraphics[width=8cm]{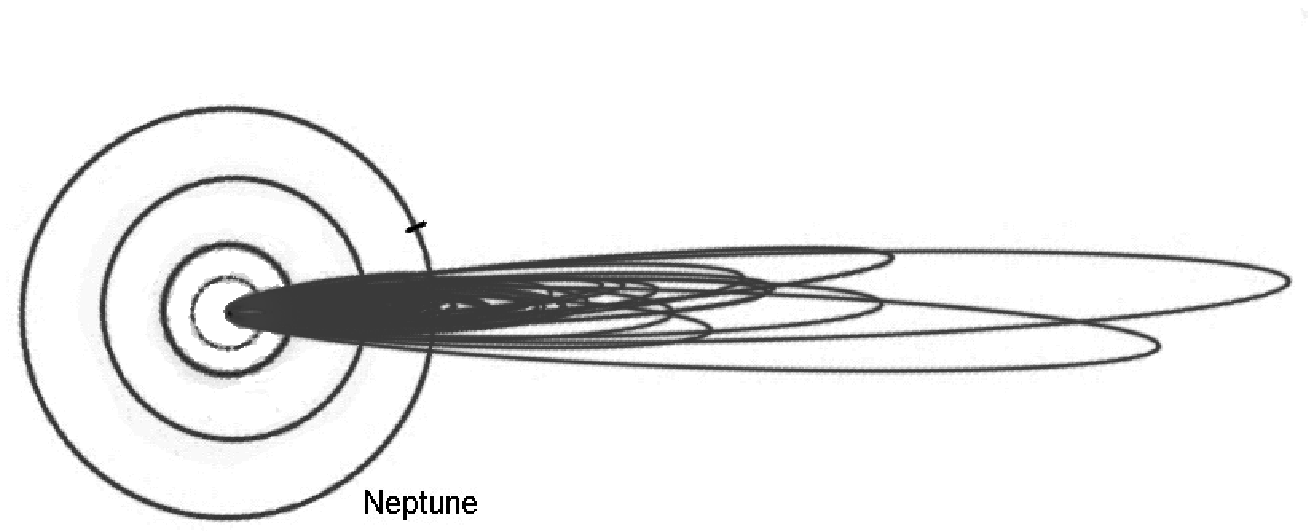}}

             \label{F1}
  \caption{Orbits of December Monocerotids and November Orionids selected from the SonotaCo database of video orbits.}
\end{figure}

\begin{figure*}
\centerline{\includegraphics[width=7.2cm]{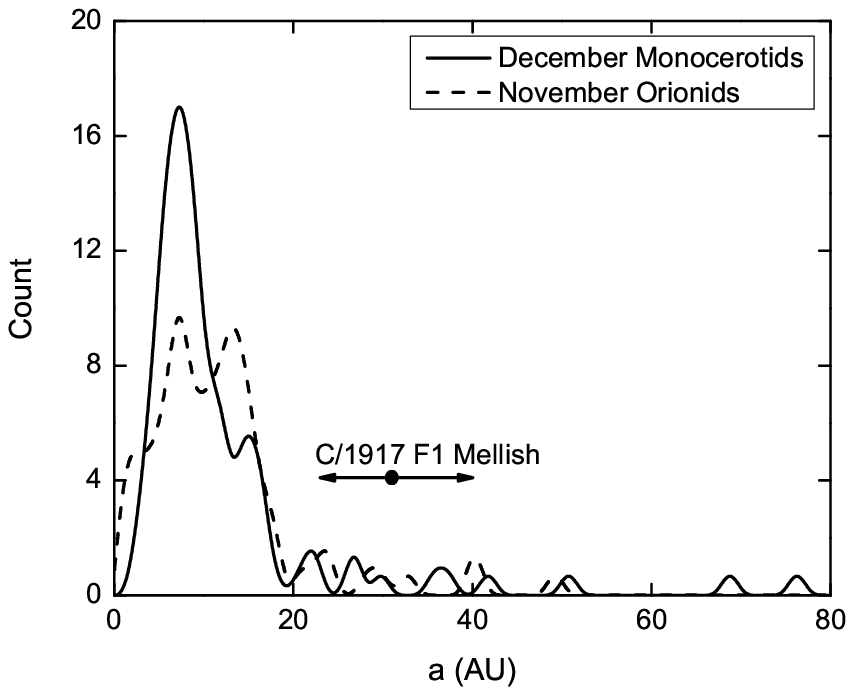}
    \hspace{0.0cm}
            \includegraphics[width=7.2cm]{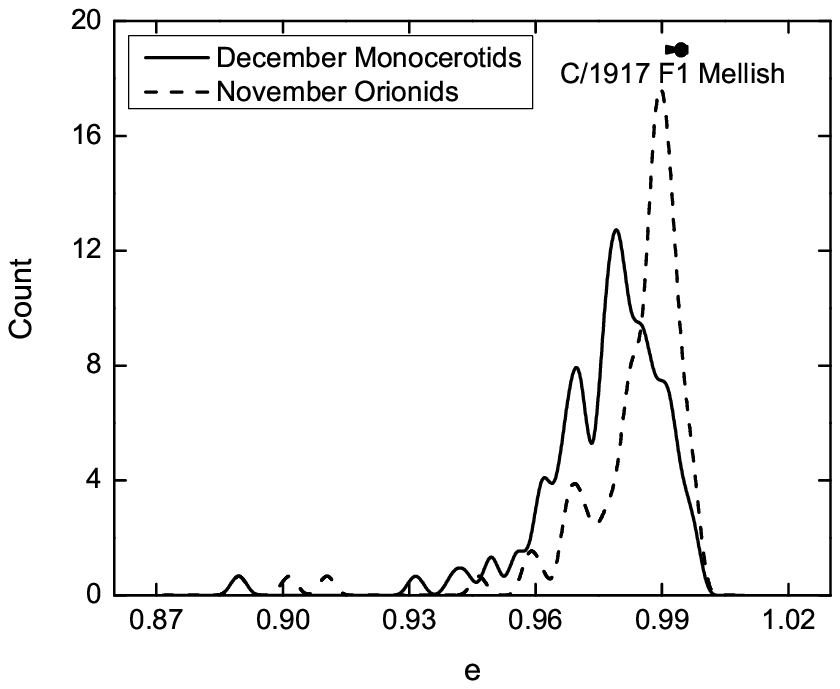}}
\centerline{\includegraphics[width=7.2cm,angle=0]{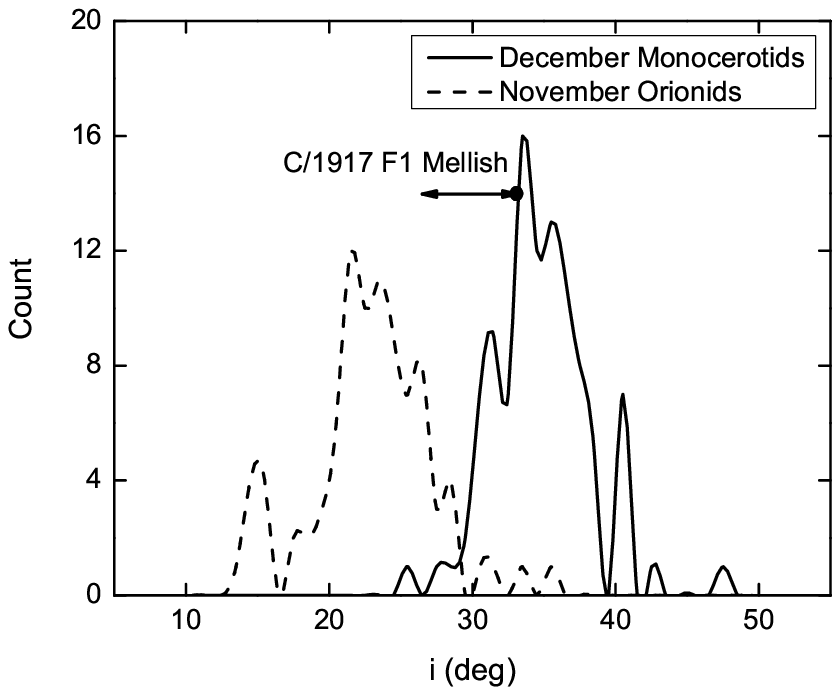}
    \hspace{0.0cm}
            \includegraphics[width=7.2cm,angle=0]{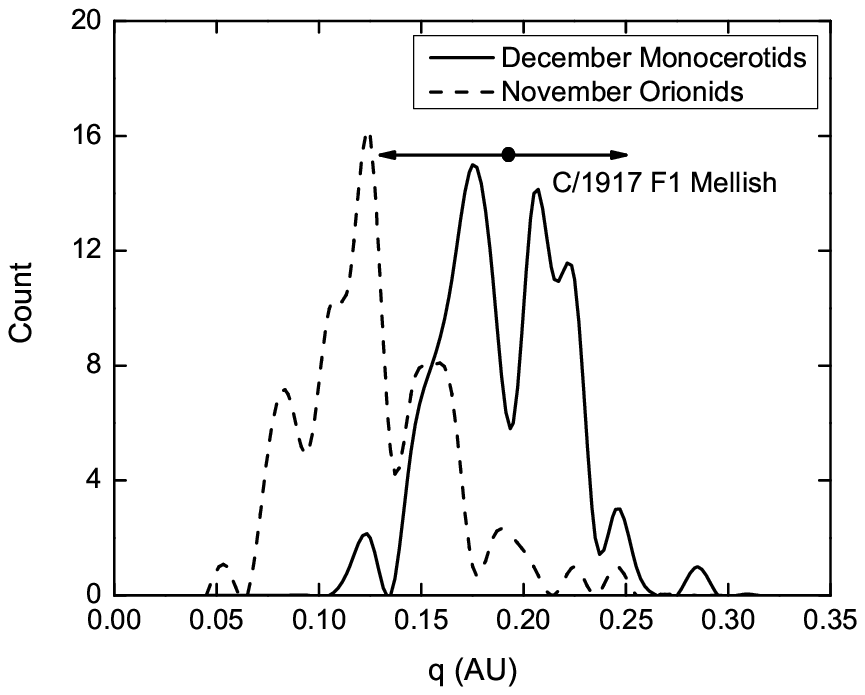}}
\centerline{\includegraphics[width=7.2cm]{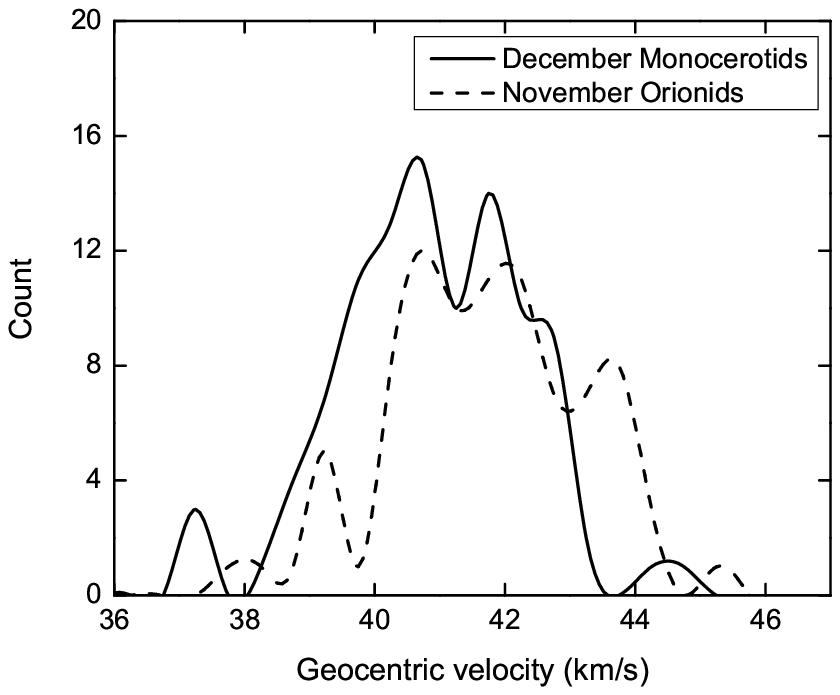}
    \hspace{0.0cm}
            \includegraphics[width=7.2cm]{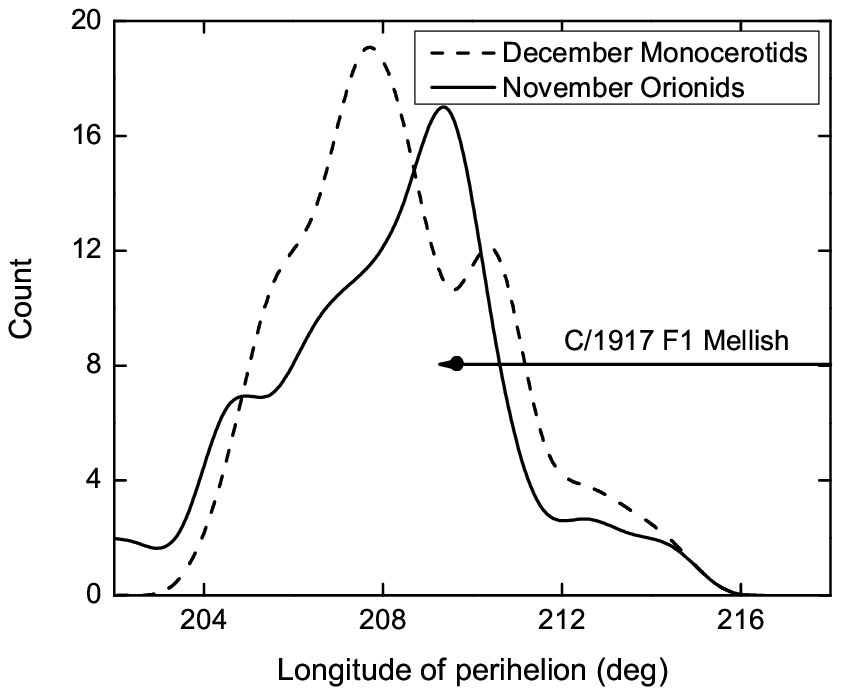}}
             \vspace{0.0cm}
  \label{F3}
\caption{Distributions of the of orbital elements and geocentric
velocity (using b--spline) of the December Monocerotids and
November Orionids. The arrow shows the orbital evolution of the
comet C/1917 F1 Mellish over the last 5000\,yr, the dot shows its
nominal orbital parameters.}
\end{figure*}

\begin{table}
\small
\begin{center}
\caption{Orbital elements of the comet C/1917 F1 Mellish in
equinox J2000 reference frame (JPL).} \label{t1}
\begin{tabular}{c|c}
\hline\hline
element & value  \\
\hline
a&27.6473325 AU \\
e&0.993121\\
q&0.190186 AU    \\
i&32.6828$^\circ$        \\
$\omega$&121.3190$^\circ$    \\
$\Omega$&88.6683$^\circ$     \\
M&0.0259325$^\circ$   \\
epoch (JD) & 2421334.0\\
heliocentric distance of Ascending Node&0.783556 AU\\
heliocentric distance of Descending Node &0.250005 AU\\

\hline\hline
\end{tabular}
\end{center}
\end{table}

\section{December Monocerotids and November Orionids from SonotaCo database}

The freely accessible database of the meteors detected by the
SonotaCo network contains $64 540$ multi-station meteor orbits
with additional parameters such as the beginning and terminal
height, absolute magnitude, equatoreal and ecliptical coordinates
of the radiant, the stream assignment, etc. Although the SonotaCo
method uses its own shower assignment algorithm, we have
demonstrated (\citealt{veres10}) that its results are consistent
with the widely-used Southworth-Hawkins D-criterion $D_{SH}$
(\citealt{south63}) for meteor stream identification. In
accordance with several restriction criteria developed in our
previous work (\citealt{veres10}), we selected a high quality
subset of orbits for further analysis. The criteria fulfilled
$111$ of $250$ December Monocerotids and $110$ of $333$ November
Orionids detected in the years 2007-2009.

Of 111 MON meteors, only 8 have hyperbolic orbits and among 110
NOO meteors, 14 have semimajor axes larger than 100\,AU or
eccentricity higher than $1.0$. We also employed $D_{SH}$ to
distinguish possible rogue sporadic meteors among MON and NOO
assigned to the showers by SonotaCo. In comparison with the
nominal orbit of the comet C/1917 F1 Mellish, all MON and NOO
meteors fall within $D_{SH}<0.4$. Even within stricter
$D_{SH}<0.15$, 105 MON and 97 NOO are found (according to SonotaCo
shower assignment). Independently, we selected the MON and NOO
shower members by using the iteration procedure according to
\citet{por94}. In the iteration for the $D_{SH}$=0.15, 105 MON and
97 NOO were identified. This result is almost identical to
SonotaCo assignment of both shower members. The showers appear
very narrow in the orbital element space. The mean orbits are
presented in Table 2, in comparison with the mean orbits by other
authors. Each orbital element was calculated as the median, with
given standard deviation. The photometric mass was computed for
each meteor according to \citet{bet99} and its mean value for the
showers is mentioned in Table 2. The orbits of the meteor showers
are given in Figure 1. Figure 2 depicts selected orbital elements
of both meteor showers. Figure 3 shows the $q-i$ phase space of
individual meteors and nominal orbit of the comet C/1917 F1
Mellish.

The semimajor axes of both shower meteors seem to be much smaller
than the nominal semimajor axis of the comet or even the range of
semimajor axes derived from the 5000\,yr orbital evolution of
cometary clones (the orbital evolution is described in chapter 3).
The eccentricity, the perihelion distance and the inclination of
MON are very close to the cometary orbit: the NOO meteors exhibit
slightly lower but still high eccentricities, notably lower
inclinations and lower perihelion distances. Generally, the orbits
of MON and NOO are very similar. In $e$ and $q$ the values are the
same within the standard deviation: a notable difference is seen
only in the inclination. The gap between the MON and NOO is also
visible in Figure 1. The $q-i$ phase space and the dispersion of
orbits from the mean orbit of each shower imply that the NOO
exhibit wider dispersion. The cometary orbit is located within MON
elements and, therefore, the NOO meteoroids seem to be older than
MON. The inclination exhibits interesting behavior. The nominal
orbit of the parent comet lies within the MON orbits, while the
NOO orbits are less inclined and are apparently well separated
from the MON orbits. There seems to be no overlapping in the
inclination between the two clumps representing the MON and NOO in
Figure 2.

\begin{table*}
\small
\begin{center}
\renewcommand{\topfraction}{0.10}
\caption{The mean orbital elements, geocentric velocity,
photometric mass of meteoroids and activity intervals of high
quality orbits of December Monocerotids and November Orionids from
the SonotaCo database, compared with other authors.  No -- number
of meteors. Authors: OH89 -- \citet{oh89}, LIND90 --
\citet{lind90}, LIND99 -- \citet{lind99}, JEN06 -- \citet{jen06}.
}\label{t2}
\begin{tabular}{c|c|c|c|c|c|c|c|c|c|c|c}
\hline\hline
elements & a & e  & q & i & $\omega$ & $\Omega$ & $v_{g}$\,[km/s] & mass\,[g] & activity & No & author \\
\hline
\textbf{MON} &8.86 & 0.979 & 0.186 & 34.7 & 129.6 & 77.9 & 41.2 & 0.6 & Nov 26 - Dec 21 & 111 & this work\\
s.dev        &     & 0.018 & 0.029 & 3.9  & 4.3   & 5.4  & 1.7  &     &                 &     &          \\
       &19.88 &0.991 &0.188 & 34.9 &128.9 &80.1 &41.6 & & & 15& OH89 \\
s.dev  & 7.49 &0.026 &0.012 & 3.1  &2.1   &2.2  &1.8  & & &   &                \\
       & 20.79 &0.990&0.187 &34.9  &128.9 &80.4 &41.82& &  Nov 27 - Dec 17 & 12 & LIND90\\
 \hline
 \textbf{NOO}& 11.36 & 0.989 &  0.125 & 23.51 &139.3 &69.3& 42.0 & 0.4 & Nov 16 - Dec 16 & 110 & this work \\
s.dev       &         &0.019 & 0.033  & 4.47  & 5.8  & 5.9 & 2.1 & & & &\\
      & 12.86   & 0.9915 &0.1093 & 24.7 & 140.4 & 67.2 & 42.6 & &
Nov 16 - Dec 7 & 38
 & LIND99\\
 & 12.7 & &0.088 & 26.9 & 145.8 & 60.0 & 43.3 & & Nov 16 - Nov 29
 & 16 & JEN06 \\
  \hline\hline
\end{tabular}
\end{center}
\end{table*}

\begin{figure}
 \includegraphics[width=8.0cm,angle=0]{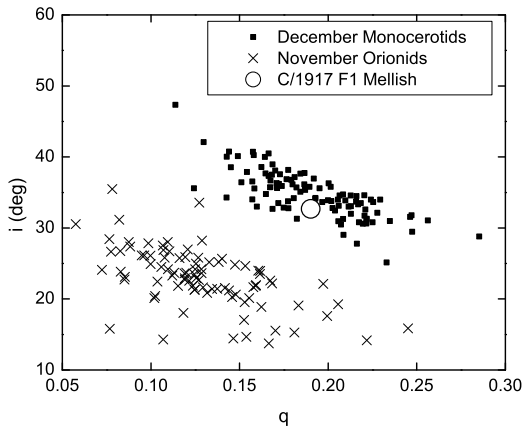}
 \caption{December Monocerotids, November Orionids and the comet C/1917 F1 Mellish in the phase space of perihelion distance
and inclination.}
  \label{F2}
\end{figure}

As seen in Table 2 and Figures 1-3, the orbital distribution of
the December Monocerotids and November Orionids have common
features. There is good agreement in the geocentric velocity and
minimal difference in the angular elements and eccentricity. The
perihelion distance of the November Orionids is generally lower.
It is worth pointing out that the perihelion distance of the MON
and NOO showers is one of the lowest observed. Only a few meteor
showers, among them the prominent Geminids, have such close
perihelion distances. The dispersion of the MON and NOO orbital
elements is relatively low, which gives the chance for good
definition and differentiation of these showers.

The cumulative absolute magnitude distribution in Figure 4 shows
that the population index of both showers is almost identical.
However, NOO contains smaller particles, according to Figure 4.
This finding is consistent with the radar data, where NOO are more
significant than MON (\citealt{nil64}, \citealt{sek73}).

We determined the maximum activity of the MON for the longitude of
the Sun $L_{\odot}=259.5$$^\circ$ (December 11) with the duration
of the shower from November 26 to December 21. The maximum of MON
occurs one day earlier in the SonotaCo data than in the IAU MDC
catalogue. The radiant position during the maximum activity was
determined as $RA=98.8$$^\circ$, $DEC=8.6$$^\circ$ and the daily
motion is given by the following equations in the equatoreal and
ecliptical coordinates:

\begin{equation}
\begin{array}{l}
 \displaystyle RA=(101.4\pm0.1)\,^\circ + (0.65\pm0.01)~(L_{\odot}-259.5\,^\circ)\\
 \displaystyle DC=(~8.0\pm0.1)\,^\circ ~~ - (0.14\pm0.02)~(L_{\odot}-259.5\,^\circ)\\
\end{array}
\label{eq:xdef}
\end{equation}

\begin{equation}
\begin{array}{l}
 \displaystyle \lambda=(101.7\pm0.1)~+(0.67\pm0.01)~(L_{\odot}-259.5\,^\circ)\\
 \displaystyle \beta=(-14.9\pm0.1)-(0.09\pm0.01)~(L_{\odot}-259.5\,^\circ),\\
\end{array}
\label{eq:xdef}
\end{equation}

\noindent where 259.5$^\circ$ represents the solar longitude of
the December Monocerotid's maximum, derived from the SonotaCo data
(eq.\,2000.0).

According to SonotaCo data, the maximum activity of the NOO occurs
on $L_{\odot}=249.5$$^\circ$ (December 1) and it is active from
November 16 until December 16. The maximum occurrs 4 days after
the maximum predicted by the IAU MDC catalogue
($L_{\odot}=245.0$$^\circ$). The motion of the radiant is given by
equations (3) and (4):

\begin{equation}
\begin{array}{l}
\displaystyle RA=(92.6\pm0.2)\,^\circ + (0.62\pm0.03)~(L_{\odot}-249.5\,^\circ)\\
 \displaystyle DC=(15.3\pm0.1)\,^\circ - (0.06\pm0.02)~(L_{\odot}-249.5\,^\circ)\\
\end{array}
\label{eq:xdef}
\end{equation}

\begin{equation}
\begin{array}{l}
 \displaystyle \lambda=(92.5\pm0.1)\,^\circ ~+ (0.60\pm0.02)~(L_{\odot}-249.5\,^\circ)\\
 \displaystyle \beta=(-8.1\pm0.1)\,^\circ - (0.06\pm0.02)~(L_{\odot}-249.5\,^\circ).\\
\end{array}
\label{eq:xdef}
\end{equation}

\noindent The sky-plane positions of the radiants in the
equatoreal and ecliptical grid are depicted in Figure 5. The
radiant positions calculated using the nominal orbit of the comet
C/1917 F1 Mellish and computed by several methods by DOSMETH
software (\citealt{nes98}) are shown as well.

\begin{figure}
 \includegraphics[width=8.0cm,angle=0]{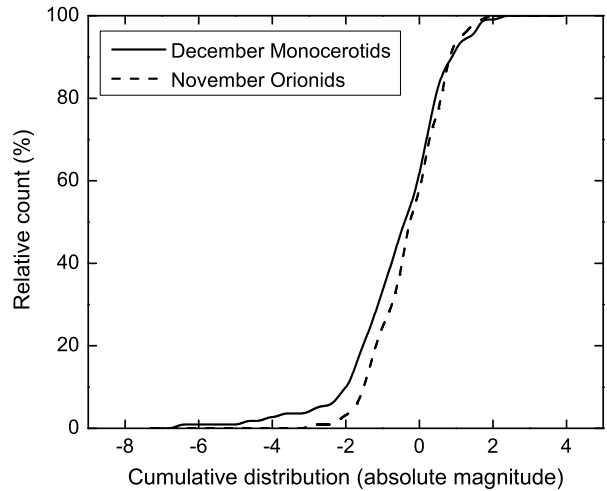}
 \caption{Cumulative distribution of absolute magnitudes of December Monocerotids and November Orionids.}
  \label{F4}
\end{figure}


\begin{figure}
 \includegraphics[width=8.0cm,angle=0]{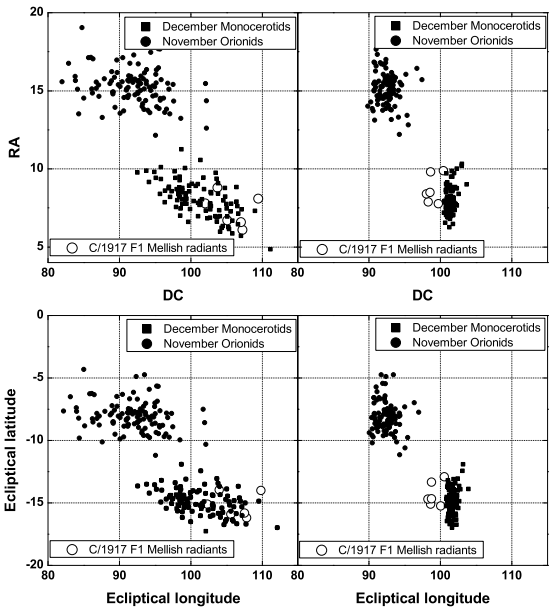}
 \caption{Equatoreal and ecliptical radiants (Eq. 2000.0) of December Monocerotids and November Orionids.
 Upper and lower right graphs show the radiants corrected for
 daily motion and reduced to the solar longitude of the activity
 maximum. Theoretical geocentric radiants derived by DOSMETH from the nominal orbit of C/1917 F1 Mellish are displayed as well (open circles).}
  \label{F5}
\end{figure}

A possible common origin of both meteor showers can also be
inferred from Figure 6 depicting the heliocentric distance of the
ascending and descending nodes of the MON and NOO as a function of
solar longitude. As expected, the ascending node lies very close
to the value of 1\,AU but the descending node gradually rises with
the solar longitude and both meteor showers overlap without any
gap or visible difference in the descending node.

Other common properties might be derived from the beginning and
terminal heights of individual meteors. Figure 7 shows clearly
that heights of the MON and NOO are practically identical. The
geocentric velocity and entry geometry is almost the same for both
meteor showers; therefore, the beginning and terminal height would
mostly depend on the physical properties of meteoroids, such as
the mass, the bulk density and internal structure. The heights are
given in Table 3 and the heights as a function of photometric mass
in Figure 7. In Table, 3 we also compare the heights of MON and
NOO with the high quality Geminids orbits from the SonotaCo
database. Geminids from the SonotaCo database have beginning and
terminal heights 5\,km lower than other video observations made by
similar techniques (\citealt{koten04}). MON and NOO meteors have
beginning heights 6--7\,km higher than Geminids, which could
indicate that meteors from C/1917 F1 Mellish have lower bulk
densities and are more fragile. On the other, hand Geminids with
similar geocentric velocities belong to the densest and most rigid
meteors observed (\citealt{ren04}).

\begin{table}
\small
\begin{center}
\caption{The beginning and terminal heights of the December
Monocerotids and November Orionids in comparison with the Geminids
using the high quality SonotaCo data.} \label{t3}
\begin{tabular}{c|c|c}
\hline\hline
shower & beginning height\,[km] & terminal height\,[km]  \\
\hline
 MON & $102.6\pm2.7$ & $86.1\pm4.8$ \\
NOO & $101.5\pm3.9$ & $85.1\pm5.3$ \\
\hline
 GEM & $95.6\pm3.3$ & $80.2\pm7.0$\\
\hline\hline
\end{tabular}
\end{center}
\end{table}

\begin{figure}
 \includegraphics[width=8.0cm,angle=0]{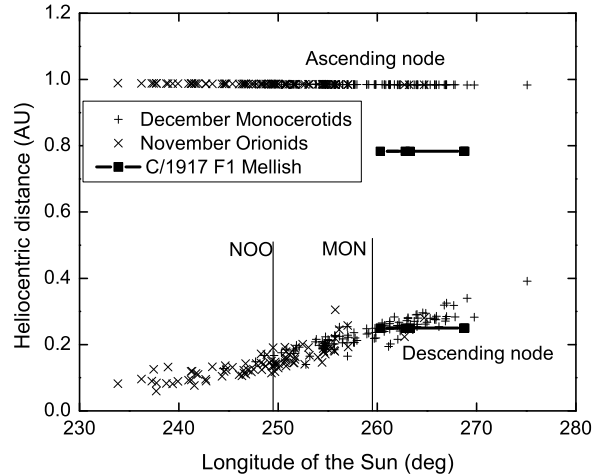}
 \caption{The heliocentric distances of the ascending and descending nodes of the December Monocerotids
 and November Orionids as a function of the longitude of the Sun,
 compared with the ascending and descending nodes of the comet C/1917 F1
 Mellish. Range of the theoretical radiants computed by DOSMETH for the parent comet is displayed with a heavy line. Vertical lines represent the maxima of the meteor showers.}
  \label{F6}
\end{figure}

\begin{table*}
\small
\begin{center}
\caption{The mean orbital elements of 6 December Canis Minorids
candidates. With respect to the parent comet, the subset has
relatively low D-criterion, $D_{SH}\sim0.33\pm0.08$} \label{t5}
\begin{tabular}{c|c|c|c|c|c|c|c|c|c}
\hline\hline
a [AU] & e & q [AU] & i & $\omega$ & $\Omega$  & $v_{g}$ & $L_{\odot}$ & RA & DC\\
$1.8\pm1.5$ & $0.95\pm0.02$ & $0.11\pm0.02$ & $30.2\pm7$ & $144.7\pm4.9$ & $78.6\pm3.1$ & $39.7\pm2.4$ & 254.1-261.6 & $107\pm4$ & $14\pm2$ \\
 \hline\hline
\end{tabular}
\end{center}
\end{table*}


\begin{figure}
 \includegraphics[width=8.0cm,angle=0]{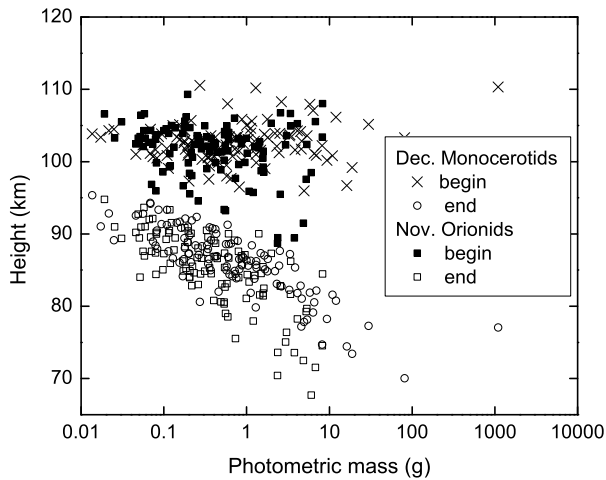}
 \caption{The beginning and terminal heights of the December Monocerotids and November Orionids as a function of the photometric mass.}
  \label{F7}
\end{figure}

The database does not contain any December Canis Minorids, yet we
tried to find some representatives among the high quality data
set. The published mean orbit of December Canis Minorids has a
very low semimajor axis, $\omega$ similar to a previously
published value, and $i$ and $\Omega$ similar to MON. According to
SonotaCo \citetext{http://sonotaco.jp/doc/J5/index.html}, the
shower might be active from November 30 until December 9, with the
maximum on December 4. Nevertheless, the database does not contain
any meteors of this stream. Because of the little information
there is about the stream, we tried to select candidates from the
high quality subset of orbits. The subset was selected using the
iteration method \citep{por94} with respect to the IAU MDC
catalogue shower parameters. Only 6 meteors fulfilled our criteria
(Table 4). Meteors were detected during the activity of both MON
and NOO. The D-criterion of meteors was on average greater than
0.3 with respect to the assumed parent comet. It is even doubtful
if the December Canis Minorids is a regular shower or if it is
only an occasional shower observed when the Earth crosses a narrow
filament of the meteoroid particles, or these meteors are just
scattered meteors of the MON-NOO complex; or that even these
meteors belong to the sporadic background.

\section{Orbital evolution of the comet and meteoroids}

The first orbital evolution analysis of the comet C/1917 F1
Mellish 800\,yr to the past \citep{car84} revealed that its orbit
evolves slowly: notably, the nodes evolve very slowly. The
inclined orbit avoids giant planet encounters and there is only a
little chance of substantial gravitational interaction with the
terrestrial planets near the perihelion. \citet{fox85} and
\citet{has99} studied the option that the ancient fireballs
observed between December 6 and 18 apparently emanated from the
same radiant. They worked out that these bolides could not be
connected to the Geminid meteor stream because of its rapid
orbital evolution but might belong to the MON. \citet{fox85}
confirmed the slow evolution of the ascending node in the 2400\,yr
integration of the cometary orbit to the past. Their work also
confirmed that the heliocentric distance of the MON ascending node
is stable as well and is close to 1\,AU one thousand years to the
past or to the future.

\begin{figure}
 \centerline{\includegraphics[width=8.0cm,angle=0]{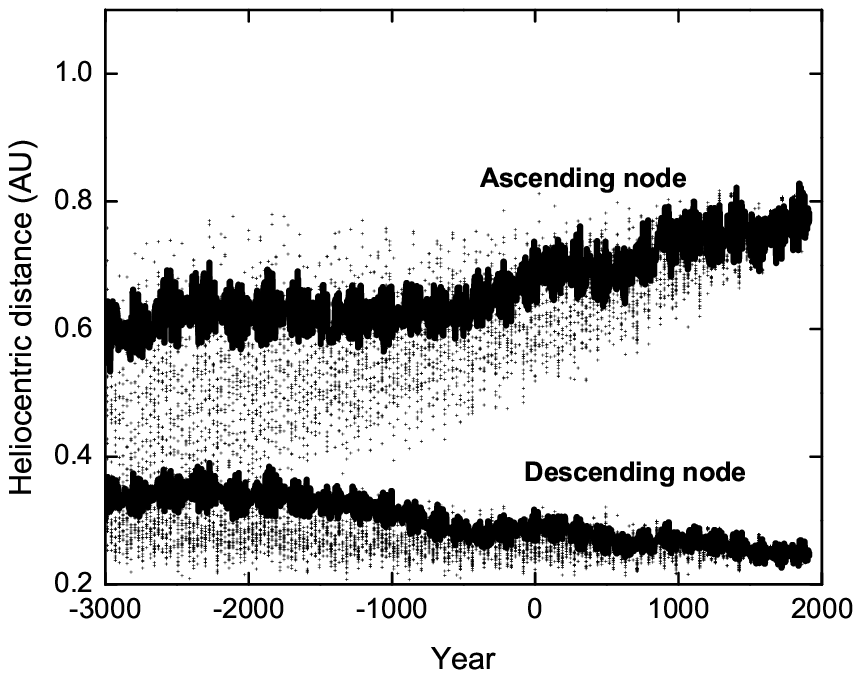}}
 \vspace{0.5cm}
 \centerline{ \includegraphics[width=8.0cm,angle=0]{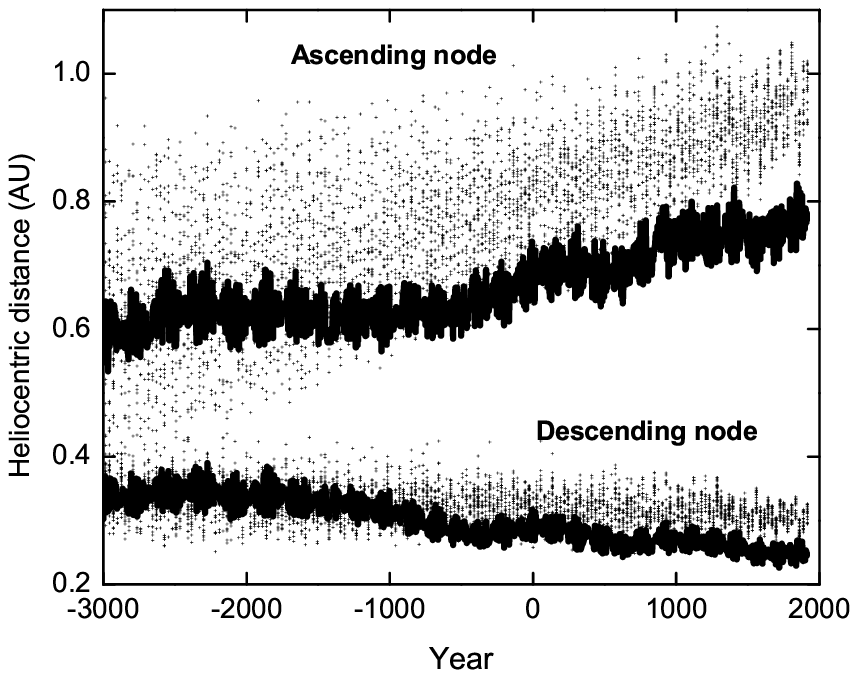}}
 \caption{The orbital evolution of the heliocentric distance of the ascending and descending nodes
 of the nominal orbit of the comet C/1917 F1 Mellish (solid line) and its clones (points).
 Up - clones generated within the 0.8\,yr orbital period error, down - clones altered in order to
put ascending node close to 1\,AU at the present day.}
  \label{F8}
\end{figure}

\begin{figure*}
\centerline{\includegraphics[width=7.0cm,angle=0]{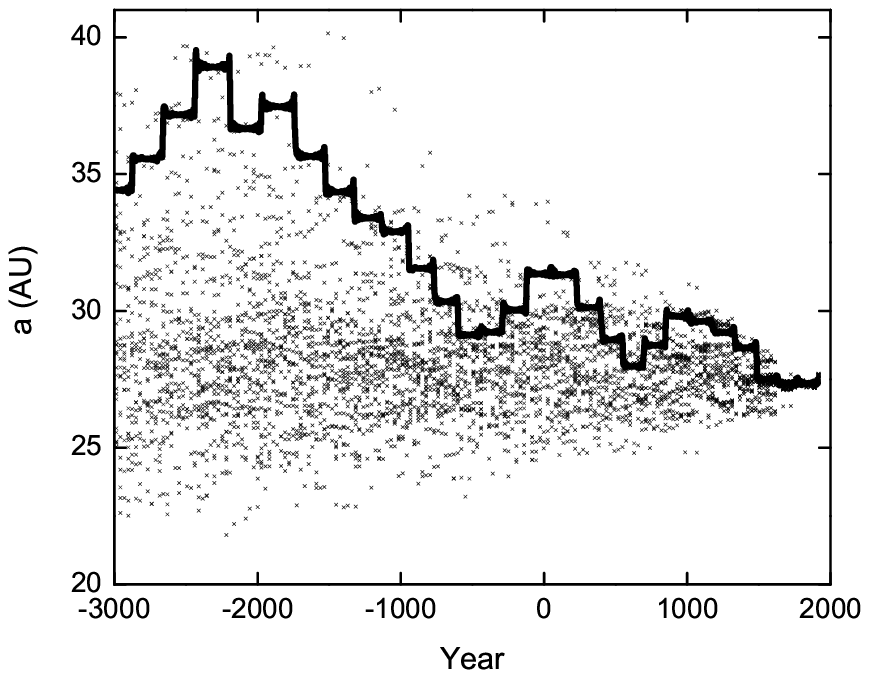}
    \hspace{0.0cm}
            \includegraphics[width=7.0cm,angle=0]{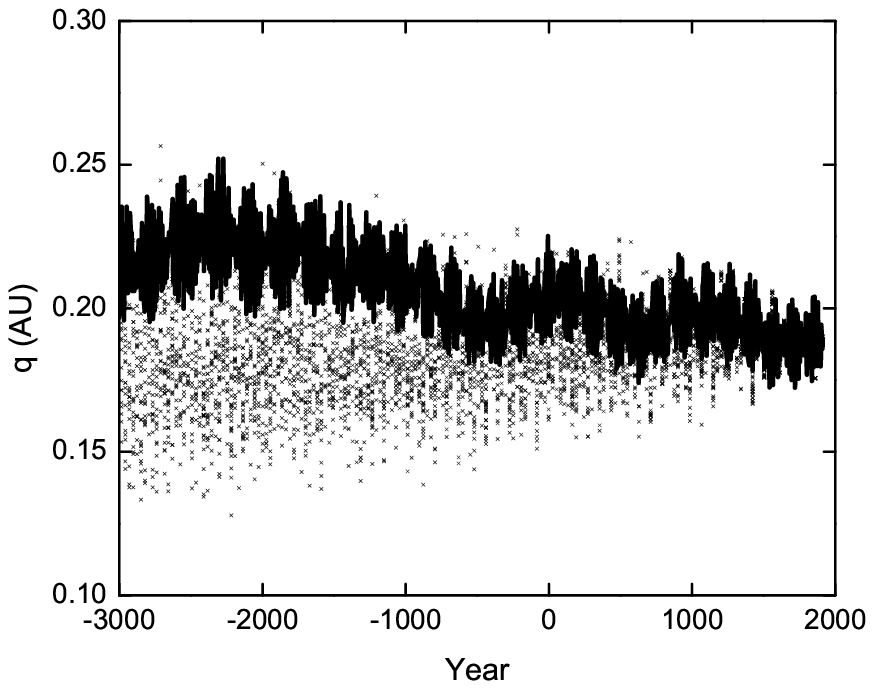}}
 \vspace{0.0cm}
\centerline{\includegraphics[width=7.0cm,angle=0]{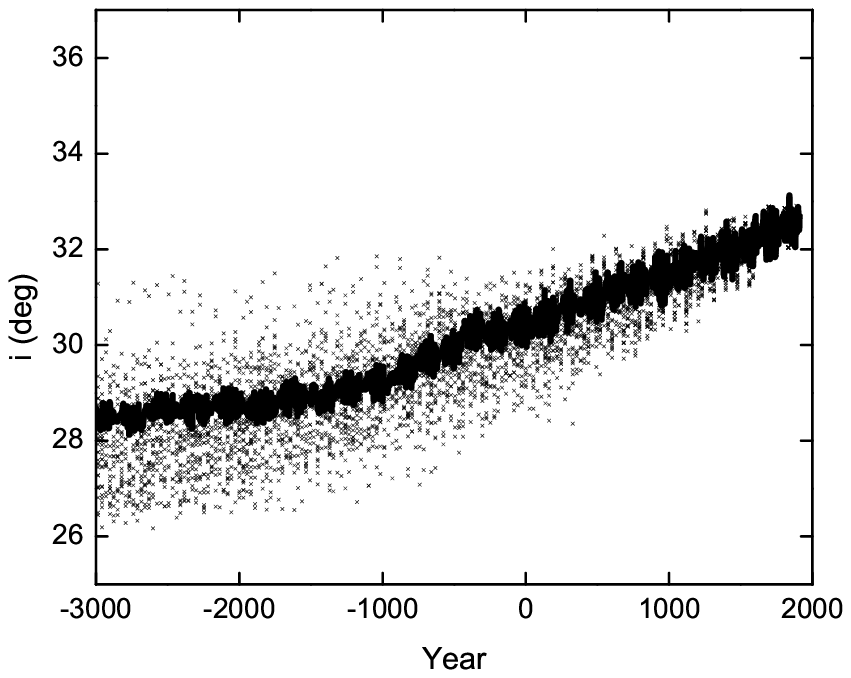}
    \hspace{0.0cm}
            \includegraphics[width=7.0cm,angle=0]{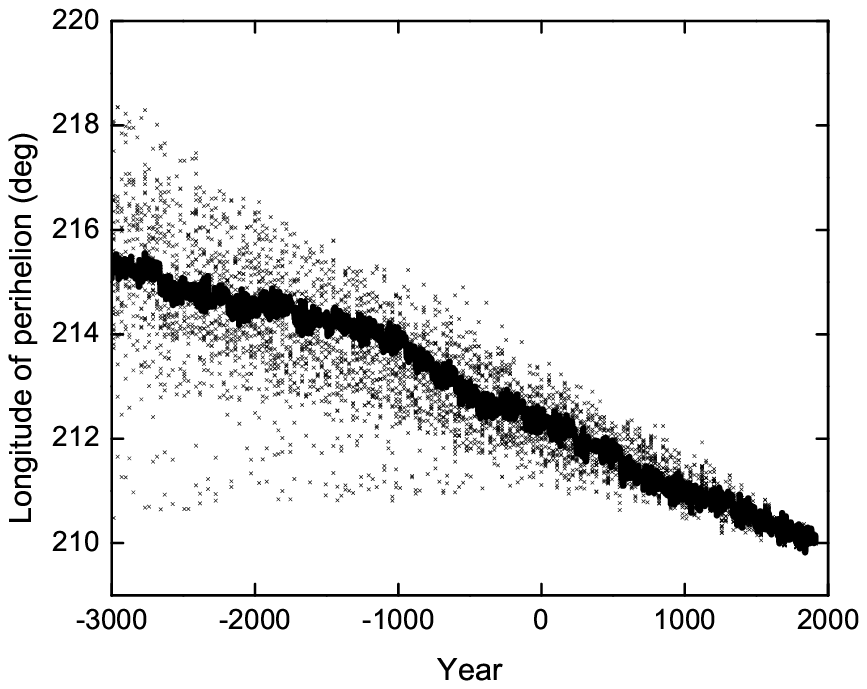}}
   \label{F9}
  \caption{The orbital evolution of clones (grey dots) and the nominal orbit of the comet C/1917 F1 Mellish (solid line).}
\end{figure*}

\begin{figure*}
\centerline{\includegraphics[width=7.0cm,angle=0]{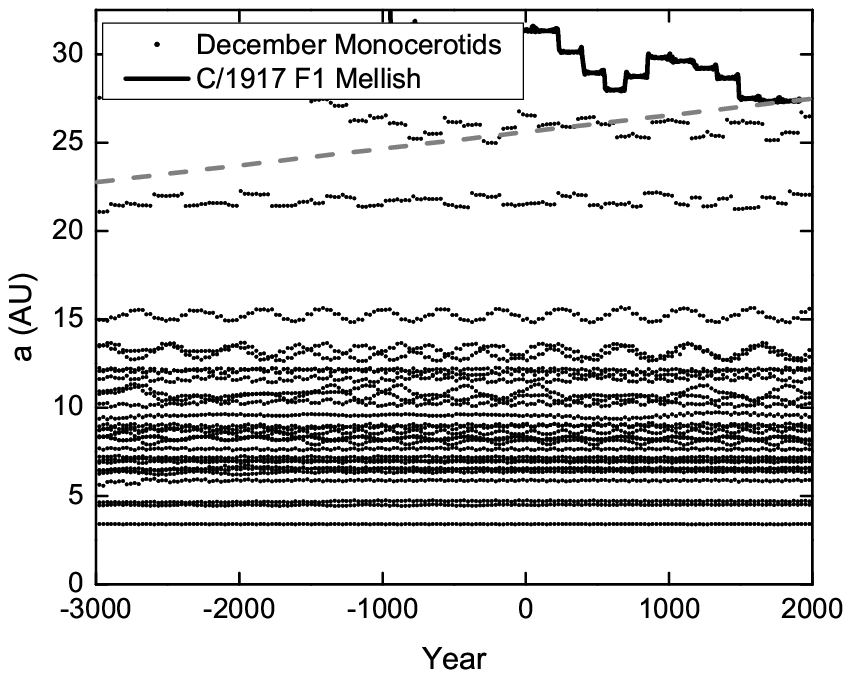}
    \hspace{0.0cm}
            \includegraphics[width=7.0cm,angle=0]{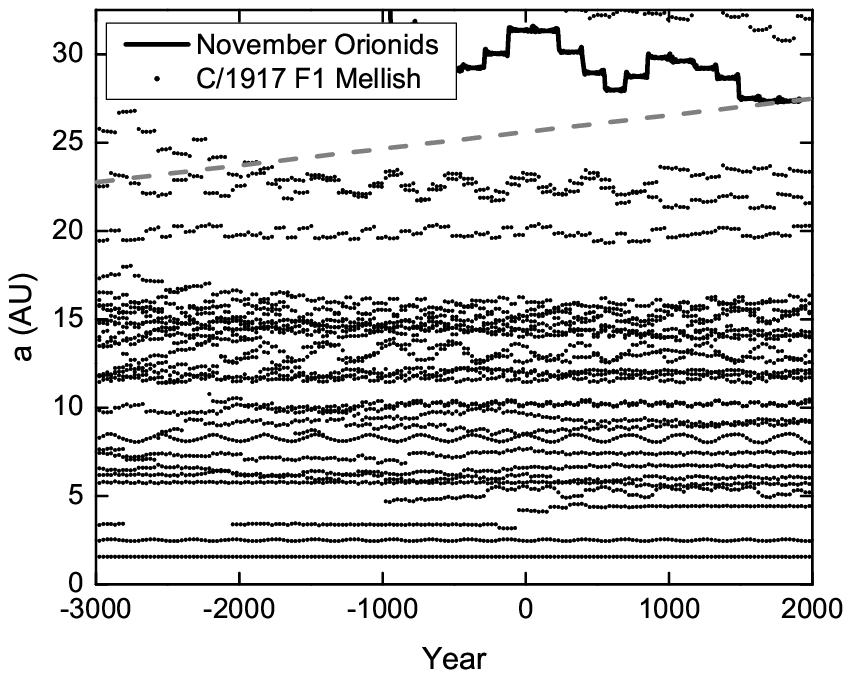}}
 \vspace{0.0cm}
\centerline{\includegraphics[width=7.0cm,angle=0]{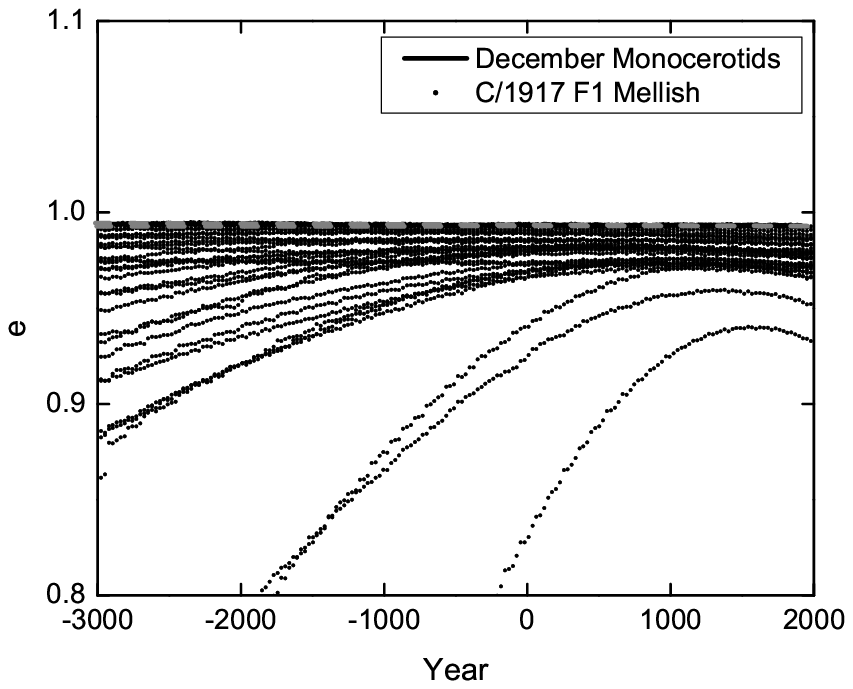}
    \hspace{0.0cm}
            \includegraphics[width=7.0cm,angle=0]{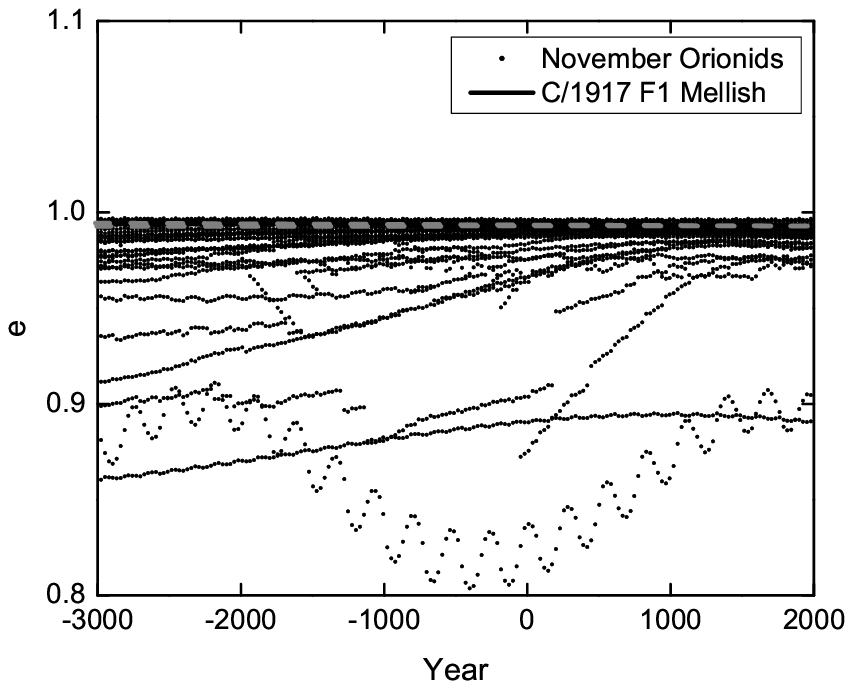}}
 \vspace{0.0cm}
\centerline{\includegraphics[width=7.0cm,angle=0]{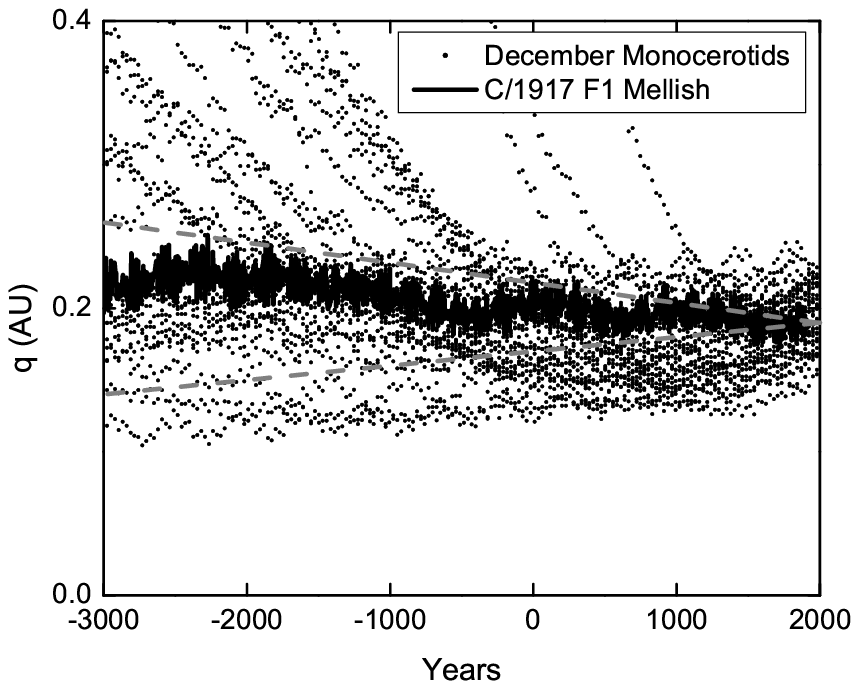}
    \hspace{0.0cm}
            \includegraphics[width=7.0cm,angle=0]{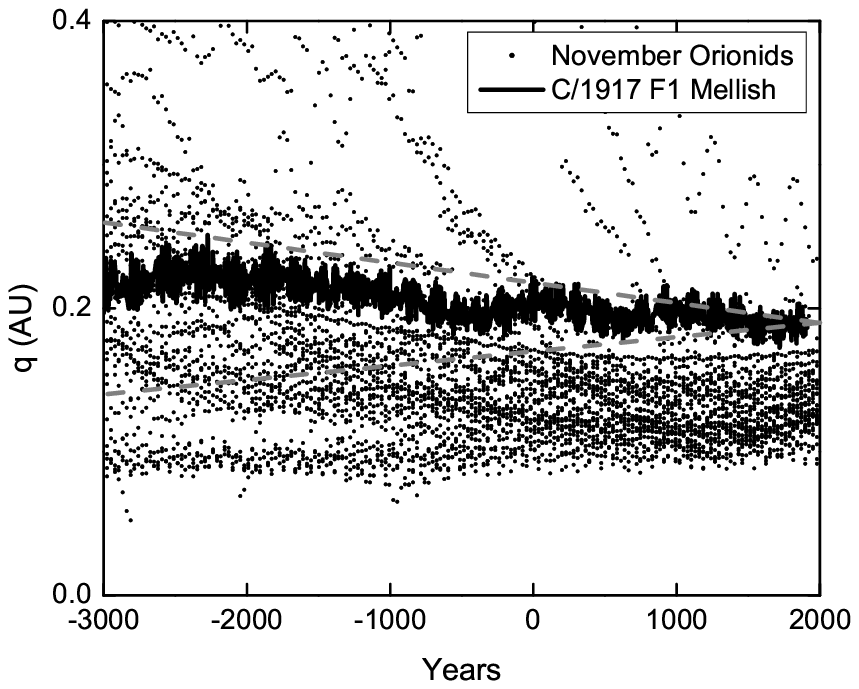}}
  \vspace{0.0cm}
\centerline{\includegraphics[width=7.0cm,angle=0]{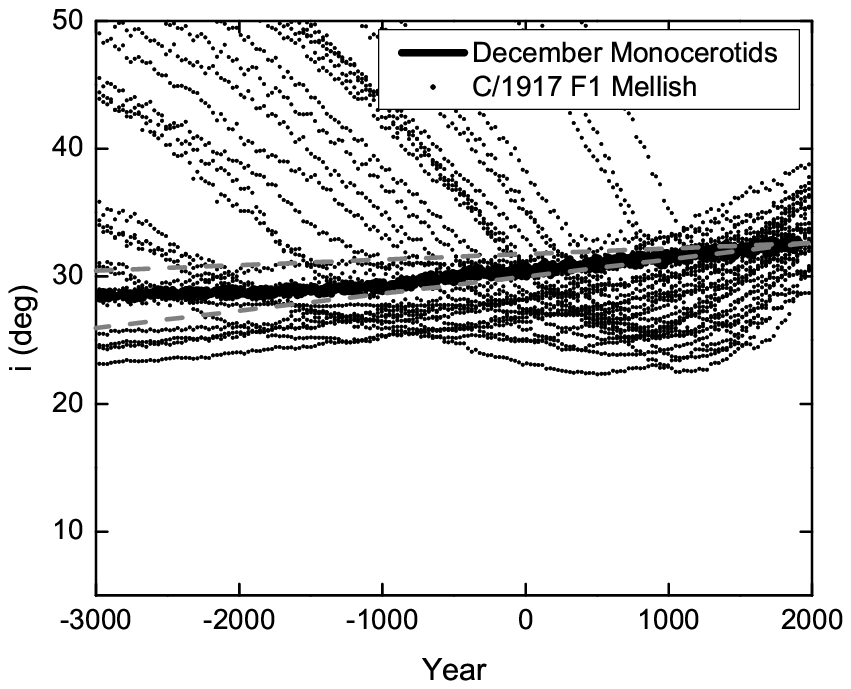}
    \hspace{0.0cm}
            \includegraphics[width=7.0cm,angle=0]{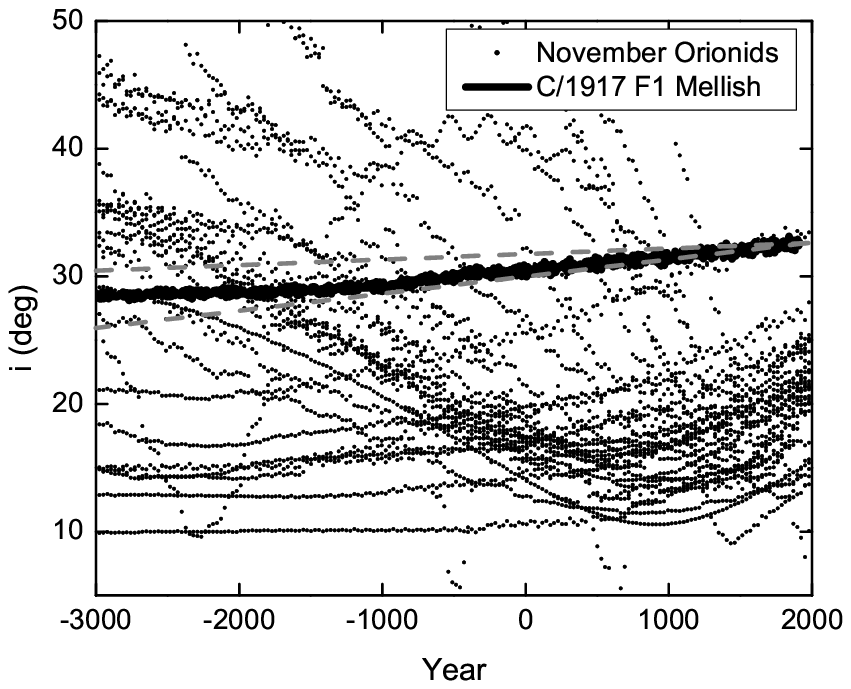}}
   \label{F10}
  \caption{The orbital evolution in $a$, $e$, $q$ and $i$ of the December Monocerotids, the November Orionids and the nominal orbit of the comet C/1917 F1 Mellish, with the possible variation shown with grey dashed lines, derived from the comet clones orbital evolution.}
\end{figure*}

\begin{figure*}
\centerline{\includegraphics[width=7.0cm,angle=0]{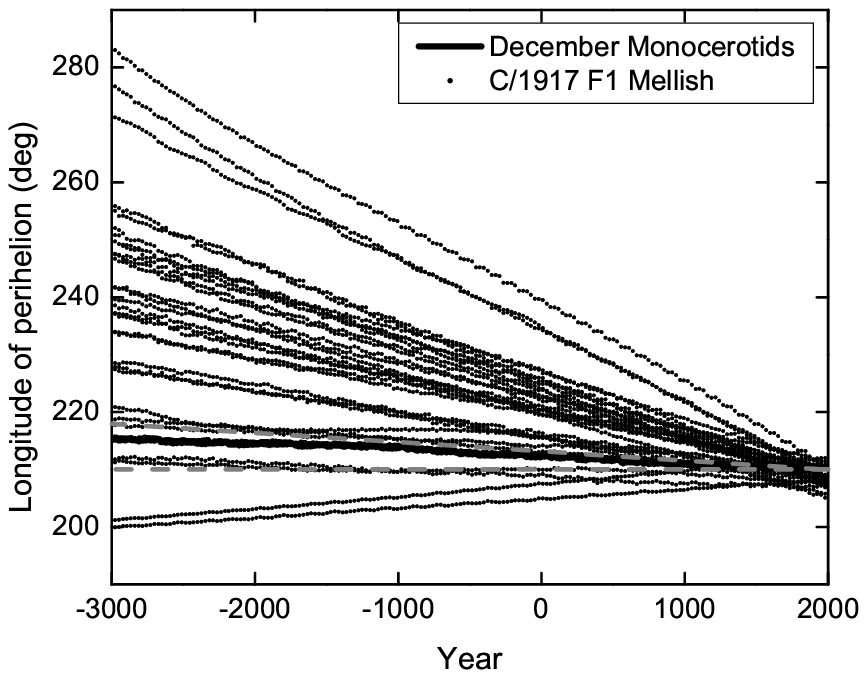}
    \hspace{0.0cm}
            \includegraphics[width=7.0cm,angle=0]{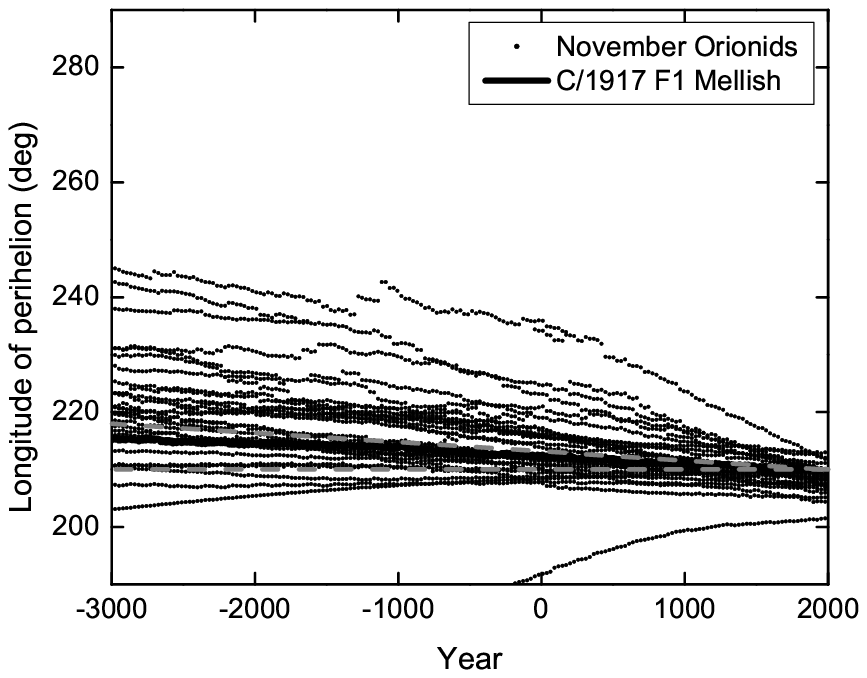}}
 \vspace{0.0cm}
\centerline{\includegraphics[width=7.0cm,angle=0]{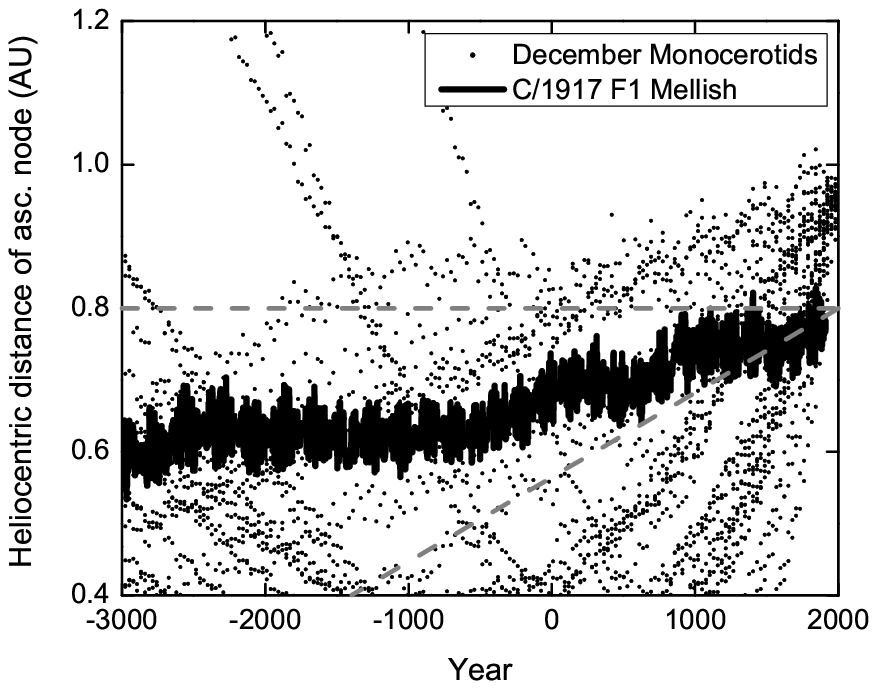}
    \hspace{0.0cm}
            \includegraphics[width=7.0cm,angle=0]{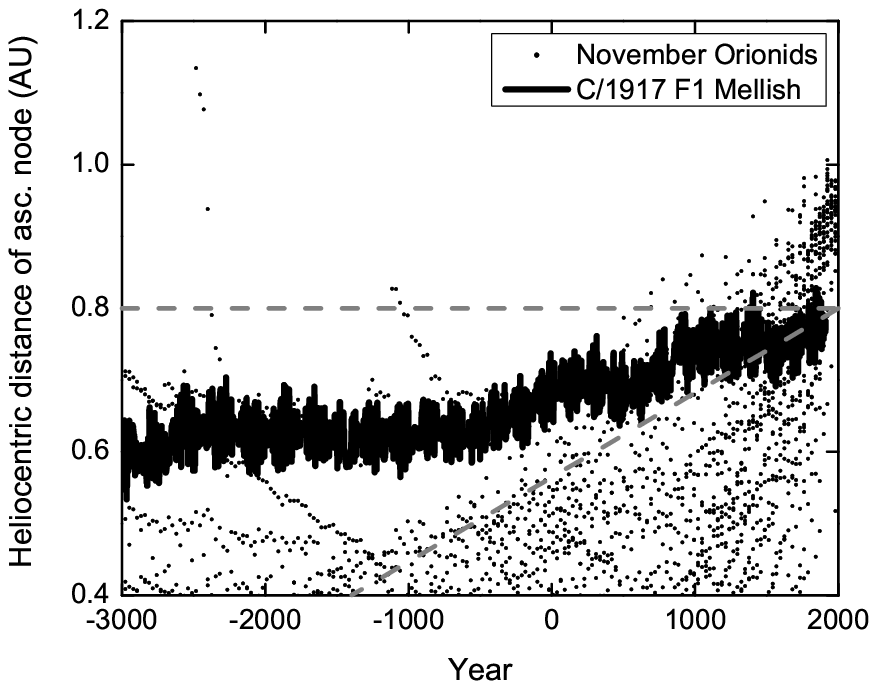}}
 \vspace{0.0cm}
\centerline{\includegraphics[width=7.0cm,angle=0]{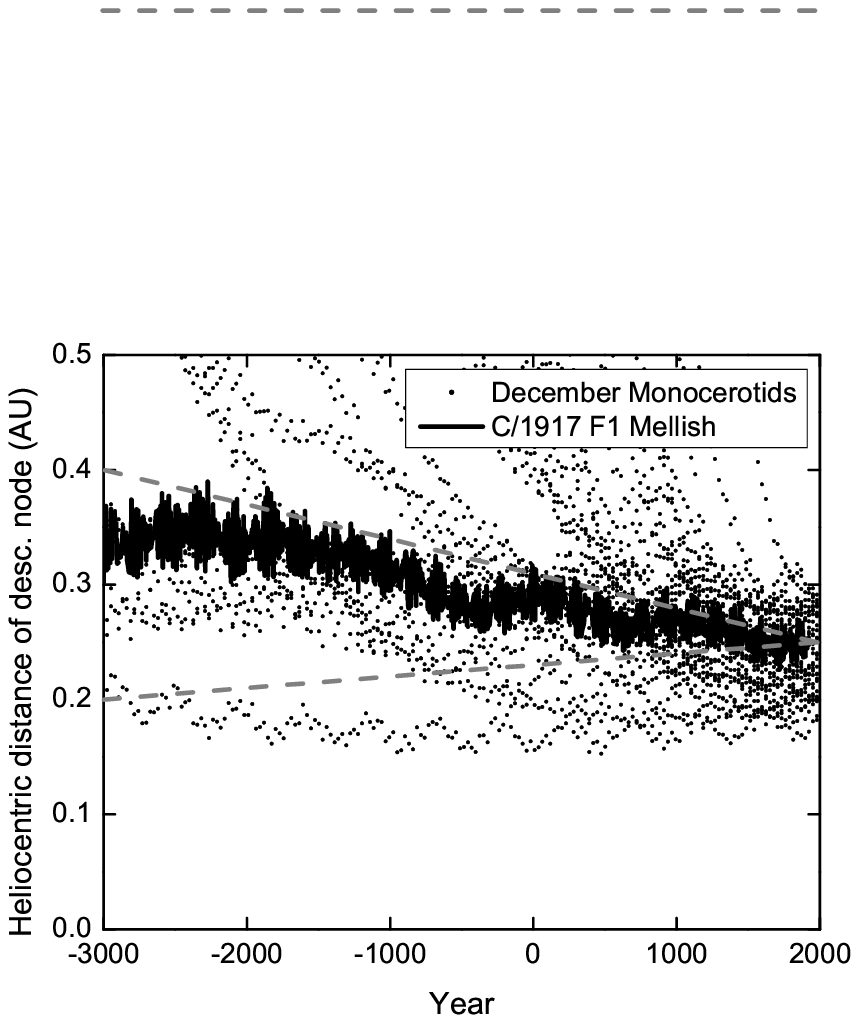}
    \hspace{0.0cm}
            \includegraphics[width=7.0cm,angle=0]{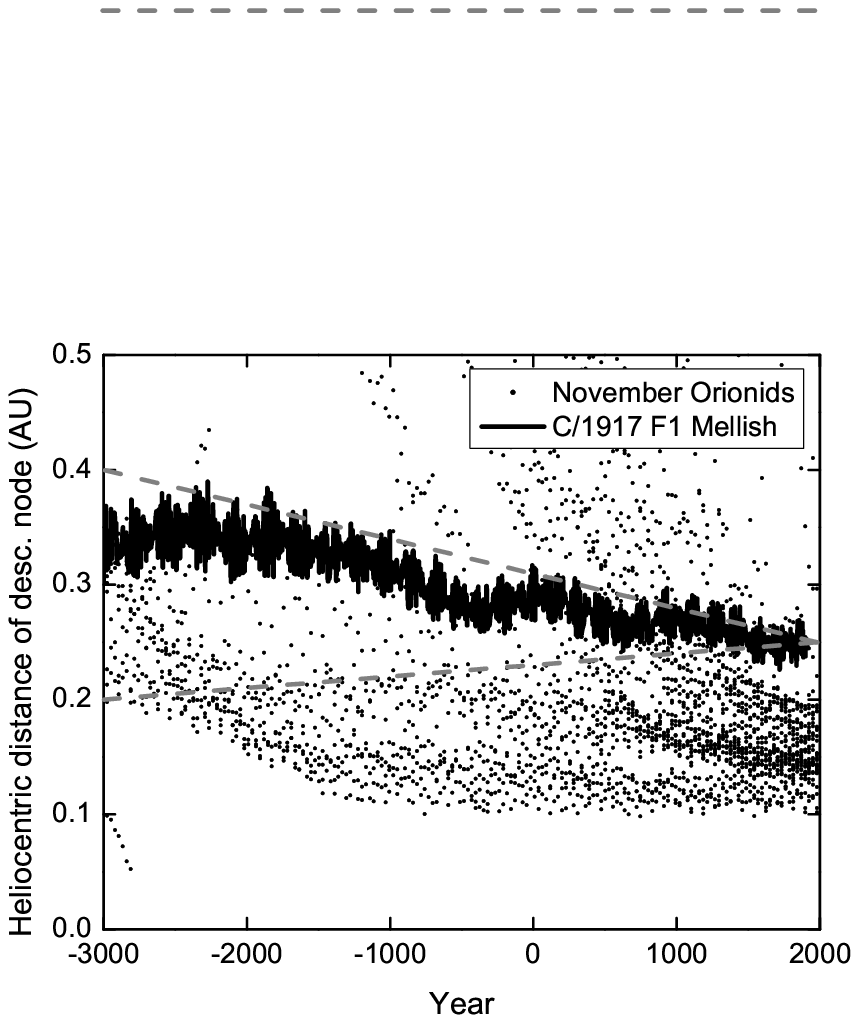}}
   \label{F11}
  \caption{The orbital evolution in $\pi$, the heliocentric distance of the ascending and descending nodes of the December Monocerotids, the November Orionids and the nominal orbit of the comet C/1917 F1 Mellish, with the possible variation shown with grey dashed lines, derived from the clones orbital evolution.}
\end{figure*}

As mentioned previously, the orbit of the parent comet was
determined with a low precision. In our study, we set out to
calculate the orbital evolution of nominal and cloned orbits of
the parent comet. We created 100 clones within the 0.8\,yr
uncertainty of the orbital period \citep{ask32}, with fixed
perihelion distance and altered semimajor axis and eccentricity
accordingly. We could not modify other orbital elements because
their uncertainties are unknown. Another set of 100 clones was
made in order to create orbits with the heliocentric distance of
ascending node close to 1\,AU while neither nominal orbit nor
first 100 clones have the ascending node close enough to the orbit
of the Earth to create an observed meteor shower. In this case,
the eccentricity, the semimajor axis and the perihelion distance
were altered. The beginning of the integration was set at the
epoch of the perihelion passage of C/1917 F1 Mellish (JD
2421334.0, Eq. 2000.0). The multistep Adams-Bashforth-Moulton type
up to 12th order numerical integrator, with variable step-width,
was used. All planets were considered as perturbing bodies and the
Earth and Moon were treated separately.

The integration shows that both sets of clones behave in a similar
way. Figure 8 depicts the heliocentric distance of the ascending
and descending node of the nominal orbit and the clones integrated
5000 years to the past. The ascending node of the nominal orbit
gradually falls and retreats from the orbit of the Earth to
$\sim0.4-0.6$\,AU from the Sun. The descending node is initially
close to the Sun ($\sim0.25$\,AU) and rises slowly to
($\sim0.3\pm0.1$\,AU). Clones of the altered orbit, that start
with the ascending node at the orbit of the Earth, behave the same
way but remain within 0.05\,AU of the Earth's orbit about 2000\,yr
to the past.

The evolution of the orbital elements of the comet and its clones
is depicted in Figure 9. The most probable value (1\,$\sigma$) of
the semimajor axes of cloned orbits is in the range of
$a\subset(25;30)$\,AU, while the nominal semimajor axis rises up
to 40\,AU 2500 BC. The inclination of both clones and the nominal
orbit falls from initial 32$^\circ$ down to
$i\subset(26;30)$$^\circ$. Even at 3000\,yr BC, the inclination is
not low enough to explain the low inclination of the NOO
($i\sim23.5$). The perihelion distance remains in the small
heliocentric distances ($q\subset(0.14;0.23)$\,AU) and the
summation of angular elements (longitude of perihelion -- $\pi$)
rises gradually from 210$^\circ$ to $\pi\subset(212;218)^\circ$
after 5000\,yr integration to the past. Clones derived from the
uncertain orbital period did not encounter the Earth but had close
flybies within the Hill sphere of Venus ($2\%$) and Mercury
($0.5\%$). This gives a non-zero chance that comet C/1917 F1
Mellish might encounter Venus or Mercury. In fact, this could
cause a sudden change in its inclination at least within 5000\,yr
in the past.

We also studied the orbital evolution of the MON and NOO meteors
from the SonotaCo database. For a chosen subset of high quality
orbits (non-hyperbolic, $a<100$\,AU), the integration uses a beta
parameter representing the solar radiation pressure
(\citealt{kla04}) for each particle ($\beta=2\cdot10^{-5}$)
derived from an assumed low bulk density ($\varrho=750\,kg/m^3$)
and the typical photometric mass of observed meteors (0.5\,g) and
starts for the epoch and the orbital elements valid for the moment
of the meteor observation. The numerical integration computed the
orbital evolution for 5000\,yr to the past (Figure 10 and Figure
11).

Unlike \citet{fox85}, perturbed orbits of the MON and NOO reduced
their heliocentric distance of the ascending nodes as time goes to
the past. It seems that it takes only two or three hundreds years
until the ascending node reaches the nominal orbit of the parent
comet. On the contrary, the semimajor axes of both the MON and NOO
are generally constant over 5000\,yr, which might be due to low
perturbations of giant planets and higher inclinations. Even with
the beta parameter, the semimajor axes remain far away from the
nominal or even cloned orbits of the parent comet. Meteoroids
could be injected to these orbits directly after the ejection from
the cometary core. The inclination of the MON is quite consistent
with the current orbital inclination of the comet C/1917 F1
Mellish. The orbital evolution in inclination implies that MON
were released recently, generally hundreds, or at most, 3000 years
in the past. On the contrary, the NOO meteoroids have lower
inclinations in the present day. The orbital evolution reveals
that NOO meteoroids could have departed from the comet mostly 4000
years prior and almost all low inclined NOO could be explained by
an orbital evolution within the last 5000 years -- most of the NOO
inclinations intersected the comet evolution path of its
inclination. The longitudes of the perihelia of the MON attain the
same values as the nominal cometary orbit in the recent centuries
and then disperse. On the other hand, the longitudes of the
perihelia of the NOO remain much longer along the evolved nominal
orbit of the comet and disperse slowly after thousands of years,
which could support the younger age of MON as well. A similar
feature is visible for the perihelion distance of both showers.
Currently the perihelion distance of MON fits well with the
current orbit of the comet but gets more dispersed around 500 B.C.
The perihelion distance of NOO is slightly different during the
last 2000 years but generally intercepts the perihelion distance
of the comet earlier than 1000 B.C. The eccentricity of MON is
dispersed much more in the past than in the case of NOO. The
heliocentric distance of the ascending node of MON lies close to
the current ascending node of the comet but disperses fast in the
past. This distance is currently lower for NOO but could be
explained by the orbital evolution (Figure 11). These implications
of the orbital evolution suggest that the NOO shower is older than
the MON shower and both streams may originate from the same parent
comet.

Resulting from the nodal distances of the comet and  the
relatively fast evolution of the ascending nodes of the showers
(centuries), there is a possibility that we observe the outer edge
of a widely evolved complex stream. The streams (MON and NOO)
might be observed as two streams as a result of a geometric
selection effect. Meteoroids with inclinations between the MON and
NOO might have nodes on non-Earth crossing orbits.

\section{Ejection of meteoroid particles from the parent comet}

The relatively large heliocentric distance of the ascending node
of the parent comet and the much lower and stable semimajor axes
of the meteors indicate that these particles were injected into
these orbits immediately after ejection from the comet, while the
comet might not have been on the same orbit at that time. The
derivation of the ejection velocity depends on the model used. If
we assume a spherical cometary nucleus with an albedo of $0.04$,
active surface $0.15\%$ (\citealt{ma02}, bulk density
$750\,kg/m^3$, a radius of the nucleus 3.1\,km (\citealt{jen06}),
ejection at the perihelion $q\sim0.19$\,AU; and if the meteoroids
are escaping only from the Sun-facing hemisphere with the Gaussian
distribution of velocities with the center on the subsolar point,
the maximum ejection velocity might range from 5\,m/s
\citep{imp01} to 112\,m/s \citep{crifo,ma02}. The escape velocity
of the meteoroid particle changes its orbital elements so that we
may calculate according to \citet{pec97}.

On the other hand, if we know the orbital elements of the comet
(before the ejection of the meteoroid) and the meteoroid after the
ejection (assuming that the observed meteoroid escaped the comet
recently and did not undergo rapid orbital changes due to
gravitational and non-gravitational perturbations), we may
directly derive the ejection velocities from equations by
\citet{pec97} as a low estimate. The ejection velocity affects, at
most, the semimajor axis. In the perihelion, the semimajor axis
change is ruled by the transversal component of the velocity
vector $\Delta v_{t}$. The range of ejection velocities derived
according to mean, the peak and minimum-maximum values of the
semimajor axis (Figure 2) of each meteor shower, in comparison
with the range of semimajor axes of the cometary clones,
integrated 2000\,yr to the past is presented in Table 5. The mean
and peak values of the transversal ejection velocity components
are in good agreement with the ejection model by \citet{ma02}.
While the eccentricity, the inclination and angular elements of
the MON lie within the cometary clones' range, the derivation of
the radial and normal components of the velocity vector are
ambiguous, following equations by \citet{pec97}. On the other
hand, a difference in the inclination, the eccentricity and
angular elements of the NOO could not be explained by the direct
ejection of the particles into current orbits, but only by the
following orbital evolution.

\begin{table}
\small
\begin{center}
\caption{The transversal component of the meteoroid ejection
velocity derived from the the difference of the semimajor axis of
meteor showers and the parent comet. The range of cometary
semimajor axis lies within $a\subset(26;29)$\,AU.} \label{t4}
\begin{tabular}{c|c|c|c|}
\hline\hline
shower & method & a [AU] & $-\Delta\,v_{t}$ [m/s]  \\
\hline
 MON & mean & 8.8 & $104\pm8$  \\
     & peak & 6.5 & $117\pm8$  \\
     & range & 3-17 & $50-145$ \\
\hline
NOO & mean & 11.36 & $90\pm8$  \\
    & peak & 6.5   & $117\pm8$ \\
    &range & 2-17  & $50-150$ \\
\hline\hline
\end{tabular}
\end{center}
\end{table}

\section{Conclusions}
We demonstrated that the December Monocerotids and November
Orionids obtained from the SonotaCo database of 3 year
observations (2007-2009) have most likely a common origin and come
from the comet C/1917 F1 Mellish. The common origin is supported
by their similar orbital characteristics, the activity, physical
properties assumed from the beginning, and the terminal heights of
the meteors, the descending nodes of both showers as a function of
the solar longitude, and the narrow Southworth-Hawkins D-criterion
for most December Monocerotids and November Orionids with respect
to the parent comet ($D_{SH}<0.15$). Direct modeling of the stream
was not an option, while the orbit uncertainty of the comet avoids
the selection of reliable and real starting point of the numerical
integration from the past to the current date. Therefore, we
studied the orbital evolution of the nominal orbit of the comet,
cometary clones within the known orbital uncertainties, and the
orbital evolution of precise meteor orbits. The dispersion of the
orbital elements, the radiants and nominal orbit of the parent
comet, currently beyond the orbital elements, suggest that the
November Orionids is an older stream than December Monocerotids.
The orbital evolution of both streams, the nominal orbit of the
comet and its clones imply that the orbital evolution is causing
November Orionids to have 10$^\circ$ lower inclinations than
December Monocerotids. There is a non-zero chance of a close
encounter of the comet with Venus or Mercury which could cause a
sudden change in the inclination of the parent comet. Furthermore,
a close encounter with a planet might cause the tidal breakup of
the comet and create a significant release of matter. The scenario
of the cometary core disintegration might also be supported by the
extremely low perihelion distance of the parent comet and both
meteor showers.

Another option is a gradual shift in the inclination, demonstrated
in the simulation. But a change in the inclination of more than
$10\circ$ of the parent comet would be solved only through a
longer orbital evolution. This option is also obscured while we do
not observe any orbits between relatively well defined clumps of
the December Monocerotids and November Orionids in the $q-i$ phase
space and radiant sky-plane distributions. Eventually there is a
wide and massive stream of the meteoroids but only some of them
have ascending nodes close to the Earth's orbit; and due to
selectional effects, we may observe two distinguished streams and
only the distant edge of the stream. The nodal distance of the
comet is currently more than 0.2\,AU from the Earth's orbit and it
retreated in the past increasingly, as well as the ascending nodes
of the observed meteors. The observed shower meteors might have
left the cometary nucleus a few centuries ago but, due to the
stable orbit of the comet, both streams might be replenished
regularly and weak shower activity might be observed each year.
The semimajor axes of both meteor streams are much lower than the
nominal orbit of the comet or its clones' evolution 5000\,yr to
the past. Almost no change of the semimajor axes of meteoroids
within the orbital evolution suggests that these particles were
injected directly into these orbits right after ejection from the
cometary nucleus. We determined that transversal component of the
ejection velocity would be about $100 m/s$ if the ejection
occurred at the perihelion. Further precise orbits and physical
data of the December Monocerotids and November Orionids are needed
for additional research.

The key question is the accuracy of the C/1917 F1 Mellish orbit. A
new measurements of the photographic plates of the comet might
reveal a more precise orbit and bring new light onto the orbital
evolution of the comet and its meteors. Our work did not confirm
any December Canis Minorids meteors in the SonotaCo database;
however, 6 candidates were selected. Their connection to the comet
C/1917 F1 Mellish is uncertain.

\section*{Acknowledgments}

This work was supported by Slovak Grant Agency VEGA, No. 1/0636/09
and by a grant of Comenius Uviversity, No. UK/245/2010.

\end{document}